# The spin $\frac{1}{2}$ Heisenberg star with frustration II: The influence of the embedding medium

Johannes Richter, Andreas Voigt, Sven E. Krüger and Claudius Gros†

Institut für Theoretische Physik, Otto-von-Guericke-Universität Magdeburg
Postfach 4120, D-39016 Magdeburg, Germany
†Institut für Physik, Universität Dortmund, D-44221 Dortmund, Germany

**Abstract.** We investigate the spin $\frac{1}{2}$ Heisenberg star introduced in J. Richter and A. Voigt, J. Phys. A: Math. Gen. **27**, 1139 (1994) [1]. The model is defined by $H = J_1 \sum_{i=1}^{N} \mathbf{s}_0\mathbf{s}_i + J_2 H_R\{\mathbf{s}_i\}$ ; $J_1, J_2 \geq 0$ , $i = 1, ..., N$. In extension to Ref. [1] we consider a more general $H_R\{\mathbf{s}_i\}$ describing the properties of the spins surrounding the central spin $\mathbf{s}_0$. The Heisenberg star may be considered as an essential structure element of a lattice with frustration (namely a spin embedded in a magnetic matrix $H_R$) or, alternatively, as a magnetic system $H_R$ with a perturbation by an extra spin. We present some general features of the eigenvalues, the eigenfunctions as well as the spin correlation $\langle \mathbf{s}_0\mathbf{s}_i \rangle$ of the model. For $H_R$ being a linear chain, a square lattice or a Lieb-Mattis type system we present the ground state properties of the model in dependence on the frustration parameter $\alpha = J_2/J_1$. Furthermore the thermodynamic properties are calculated for $H_R$ being a Lieb–Mattis antiferromagnet.

Short title: The Heisenberg star II

## 1. Introduction

The properties of interacting quantum spins have been attracted large attention over a long period. Only some model Hamiltonians can be solved exactly. Important examples are (i) models in one dimension solvable by Bethe-Ansatz [2, 3, 4], (ii) valence-bond models [5, 6, 7], (iii) a one-dimensional model with long-range inverse-square exchange [8, 9, 10, 11], and (iv) models with long-range interaction of constant strength (Lieb-Mattis type models [12] which have been used to discuss spontaneous symmetry breaking in quantum spin systems recently [13, 14, 15]) .

In [1] (further referred as STAR I) we introduced the frustrated spin $\frac{1}{2}$ Heisenberg star with a Hamiltonian $H_I = J_1 \sum_{i=1}^{N} \mathbf{s}_0\mathbf{s}_i + J_2 \sum_{i=1}^{N} \mathbf{s}_i\mathbf{s}_{i+1}$ $(J_1, J_2 > 0)$ representing a central site with $N$ nearest neighbours which can be unconnected $(J_2 = 0)$ or connected $(J_2 \neq 0)$. The star can be considered either as an essential structure element of a lattice or as a linear chain with a perturbing extra spin.

For the above defined Hamiltonian $H_I$ we presented in STAR I general relations for the eigenvalues, the eigenstates and the spin correlation function in the ground state as well as numerical results for $N = 4, 6, ..., 22$. Analyzing the analytical and numerical data we discussed the ground state phase diagram, in particular, the ground state spin correlations versus $\alpha = J_2/J_1$. We found that for $J_2/J_1 < \alpha_{crit}$ the ground state of the system is the state with strongest antiferromagnetic correlation $\langle \mathbf{s}_0\mathbf{s}_i \rangle = -\frac{1}{4} - \frac{1}{2}\frac{1}{N}$ between the central spin $\mathbf{s}_0$ and a neighbouring spin $\mathbf{s}_i$ and with ferromagnetic correlations $\langle \mathbf{s}_i\mathbf{s}_j \rangle = \frac{1}{4}$ within the ring. If $J_2/J_1$ exceeds $\alpha_{crit}$ it follows a series of transitions to states with successively weaker correlations $\langle \mathbf{s}_0\mathbf{s}_i \rangle$ ending with $\langle \mathbf{s}_0\mathbf{s}_i \rangle = 0$ for dominating $J_2$. For $\alpha_{crit}$ we found exactly $\frac{1}{4}$ , independent of the size of the system. For larger $N$ this weakening of the antiferromagnetic correlation of the central spin takes place very rapidly when changing $J_2/J_1$ in a small region above $\alpha_{crit}$. The extrapolation $N \to \infty$ yields for $J_2/J_1 < \alpha_{crit}$ the correlator $\langle \mathbf{s}_0\mathbf{s}_i \rangle = -\frac{1}{4}$ which can be considered as an upper limit for the ground state correlation $\langle \mathbf{s}_l\mathbf{s}_m \rangle_o$ of antiferromagnetically interacting spins $l$ and $m$ in a spin $\frac{1}{2}$ Heisenberg antiferromagnet without competition between the interactions. We argued that any ground state spin correlation $\langle \mathbf{s}_i\mathbf{s}_j \rangle$ of antiferromagnetically coupled spins $\mathbf{s}_i$ and $\mathbf{s}_j$ larger than $-\frac{1}{4}$ is an effect of competing interactions. In the limit of large $J_2/J_1$ the correlation $\langle \mathbf{s}_i\mathbf{s}_j \rangle_o$ within the ring becomes antiferromagnetic and the ring state goes over to the Bethe singlet for $J_2 \approx J_1 N(\frac{2}{\pi^2} + O(\frac{1}{\ln N}))$.

In this paper we generalize the model by considering a central spin with different embedding media. In particular, we compare models where a central spin is embedded in an antiferromagnetic Heisenberg matrix of linear chain, square lattice and Lieb-Mattis type. Furthermore, we discuss in more detail the properties of the model in the thermodynamic limit $N \to \infty$. Additional to the ground state properties we present the full thermodynamics in the case of the Lieb-Mattis star.

## 2. Model and general relations

We consider a spin $\frac{1}{2}$ Heisenberg system

$$H = \frac{J_1}{N}\sum_{i=1}^{N} \mathbf{s}_0\mathbf{s}_i + J_2 H_R\{\mathbf{s}_i\} \quad ; \quad J_1, J_2 > 0 \tag{1}$$

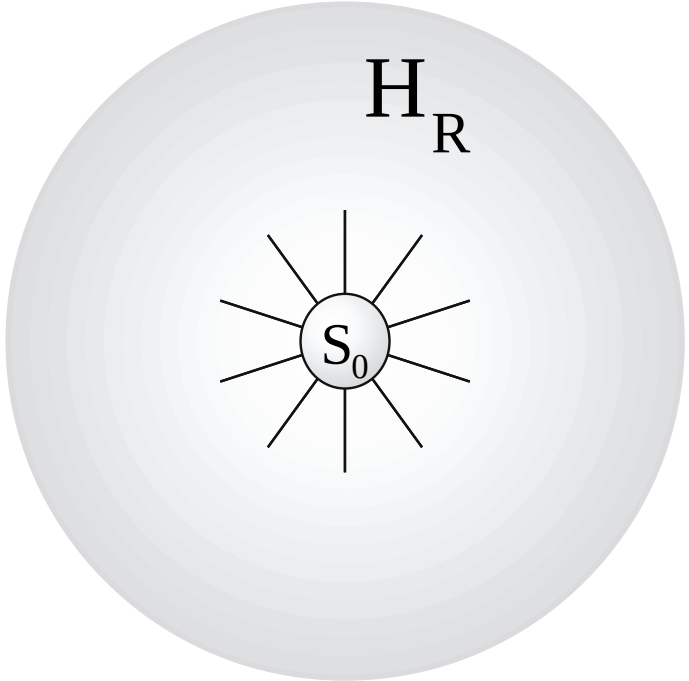

Here $H_R\{\mathbf{s}_i\}$ represents itself a spin $\frac{1}{2}$ rotationally invariant antiferromagnet describing the medium which surrounds the central spin. In (1) the interaction with the central spin was scaled by $N$. A similar scaling will also be used for $H_R$ (see below). In contrast to the paper STAR I we do not specify the Hamiltonian $H_R$ at this point.

Several general relations found for the LC star in paper STAR I are valid for the more general $H_R$, too. Let us define the total spin $\mathbf{S} = \sum_{i=0}^{N} \mathbf{s}_i$ of the system and the total spin of the embedding medium $\mathbf{S}_R = \mathbf{S} - \mathbf{s}_0 = \sum_{i=1}^{N} \mathbf{s}_i$. The following universal features of the model are important:
**(i)** The integrals of motion of the systems are $H$, $H_R$, $\mathbf{S}^2$, $S_z$, $\mathbf{S}_R^2$ with the respective quantum numbers $E$, $E_R$, $s$, $m$, $r$.
**(ii)** The eigenvalues of $H$ are given by

$$E = J_2 E_R + \frac{J_1}{2N} r \qquad \text{for} \quad s = r + \frac{1}{2} \ ; \ r = 0, 1, 2, ..., \frac{N}{2} \tag{2}$$

$$E = J_2 E_R - \frac{J_1}{2N}(r+1) \qquad \text{for} \quad s = r - \frac{1}{2} \ ; \ r = 1, 2, ..., \frac{N}{2}. \tag{3}$$

Clearly, for fixed $r$ and fixed $E_R$ the states with $r = s + \frac{1}{2}$ have lower energy. Let us look for the lowest energy in every subspace of fixed quantum number $r$. For the energy $E_R$ of the embedding antiferromagnet $H_R$ we have the Lieb-Mattis level ordering [12, 16] for the lowest eigenvalue $E_R(r)$ in the considered subspace. Hence, from the general relation (3) we indicate a competition between both energy terms leading to interesting frustration effects. In particular, at zero temperature the frustration parameter $\alpha = J_2/J_1$ determines the quantum number $r$ in the ground state and may be used to tune the ground state properties.

**(iii)** The basic structure of the eigenstates of (1) reads as follows

$$|\Phi_{E,s,m,r}\rangle = a \mid\uparrow\rangle|\Phi^R_{E_R,r,m-1/2}\rangle + b \mid\downarrow\rangle|\Phi^R_{E_R,r,m+1/2}\rangle \ , \tag{4}$$

where $\mid\uparrow\rangle$ and $\mid\downarrow\rangle$ are the eigenfunctions of the z-component of the central spin $s_{0,z}$ and the $|\Phi^R_{E_R,r,m\pm1/2}\rangle$ are the eigenfunctions of $H_R$, $\mathbf{S}^2_R$, $S_{R,z}$ with the corresponding eigenvalues $E_R$, $r(r+1)$, $m \pm 1/2$. The eigenfunctions $|\Phi^R_{E_R,r,m+1/2}\rangle$ and $|\Phi^R_{E_R,r,m-1/2}\rangle$ of $H_R$ are related to each other by

$$|\Phi^R_{E_R,r,m-1/2}\rangle = [r(r+1) - (m+1/2)(m-1/2)]^{-1/2} S^-_R |\Phi^R_{E_R,r,m+1/2}\rangle \tag{5}$$

The coefficients $a$ and $b$ do not depend on $J_1$, $J_2$ and are given by

$$a = \sqrt{\frac{r+m+\frac{1}{2}}{2r+1}} \ , \ b = \sqrt{\frac{r-m+\frac{1}{2}}{2r+1}} \quad \text{for} \quad s = r + \frac{1}{2} \tag{6}$$

$$a = -\sqrt{\frac{r-m+\frac{1}{2}}{2r+1}} \ , \ b = \sqrt{\frac{r+m+\frac{1}{2}}{2r+1}} \quad \text{for} \quad s = r - \frac{1}{2} \quad . \tag{7}$$

**(iv)** For the spin correlation of the central spin with a spin of embedding medium we have

$$\langle\Phi_{E,s,m,r}|\mathbf{s}_0\mathbf{s}_i|\Phi_{E,s,m,r}\rangle = \frac{r}{2(N-1)} \quad \text{for} \quad s = r + \frac{1}{2} \tag{8}$$

$$\langle\Phi_{E,s,m,r}|\mathbf{s}_0\mathbf{s}_i|\Phi_{E,s,m,r}\rangle = \frac{-(r+1)}{2(N-1)} \quad \text{for} \quad s = r - \frac{1}{2} \quad . \tag{9}$$

The spin correlation within the embedding medium, of course, depends on the detailed structure of $H_R$.

The more specific properties of the model are determined by properties of the embedding medium entering the general relations (2) – (9) via $E_R$, $|\Phi^R_{E_R,r,m\pm1/2}\rangle$. Whenever the solution for the Hamiltonian of the surrounding medium $H_R$ is known (i.e. $E_R$ and $|\Phi^R_{E_R,r,m\pm1/2}\rangle$ are available) the total system is solvable.

We consider as solvable systems the linear chain (LC) and the Lieb-Mattis (LM) antiferromagnet. Both model represent the extreme limits of the Heisenberg antiferromagnet with maximal (LC) and minimal (LM) quantum fluctuations. Additionally to the solvable limits we present approximative results for the square-lattice (SL) antiferromagnet. In order to compare the different cases we scale the interactions by the number of sites $N$ so that the energy becomes intensive and that the energy of the fully polarized ferromagnetic state is identical $E_R^{fm} = \frac{1}{4}$ for all embedding media considered. This scaling is consistent with the $1/N$-scaling of the interaction of the central spin (see equation (1)). We write for the Hamiltonians of the media

$$H_R^{LC} = \frac{1}{N}\sum_{i=1}^{N} \mathbf{s}_i\mathbf{s}_{i+1} \tag{10}$$

$$H_R^{SL} = \frac{1}{2N}\sum_{i=1}^{N}(\mathbf{s}_i\mathbf{s}_{i+\hat{x}} + \mathbf{s}_i\mathbf{s}_{i+\hat{y}}) \tag{11}$$

$$H_R^{LM} = \frac{4}{N^2}\sum_{\substack{i,j=1\\ i\in A, j\in B}}^{N} \mathbf{s}_i\mathbf{s}_j. \tag{12}$$

In what follows we fix $J_1 = 1$ and consider $\alpha = J_2/J_1$ as the parameter of the model.

## 3. Finite systems

For small $N$ the star can serve as an elementary cluster of a lattice with frustration. We focus our consideration in this section on the example of N=8. For that $N$ the LM system is by chance identical with the SL. For comparison we add to the discussion of the LC, SL and LM star the elementary cube of the body-centered cubic lattice (BCC). The calculation of the spectra and the wave function can be simply done by numerically exact diagonalization. The results for the relevant spin correlation of the central spin $\mathbf{s}_0$ with a neighbouring spin $\mathbf{s}_i$ are presented in figure 1. The steps in the correlation functions indicate the transitions between states with different quantum number $r$. Of course, the frustrating $J_2$ interaction weakens the antiferromagnetic correlation. But it is interesting that independent of the embedding medium an essential diminishing of the strength of the correlation takes place in a small region above $\alpha_{crit} = (J_2/J_1)_{crit} = \frac{1}{4}$. This region is in particular small if the surrounding medium has a low number of nearest neighbours (LC). Otherwise, the complete suppression of the antiferromagnetic correlation by the frustrating $J_2$ takes place for fairly large $\alpha$ (precisely at $N/4$ for LM and at slightly lower values for LC and BCC).

For small $N$ one can calculate the full thermodynamics. As an example we present the temperature dependence of $\langle \mathbf{s}_0\mathbf{s}_i\rangle$ and $\langle \mathbf{s}_i\mathbf{s}_j\rangle$ for the LC star in figure 2. (The behaviour is qualitatively the same for the other systems.) For $\alpha < \frac{1}{4}$ the behaviour is the standard one for both $\langle \mathbf{s}_0\mathbf{s}_i\rangle$ and $\langle \mathbf{s}_i\mathbf{s}_j\rangle$, i.e. the strength of the correlation is diminished by thermal fluctuations. Of course, there is no sharp transition for the finite system. For $\alpha > \frac{1}{4}$ the frustration is more important and we find for low temperatures a qualitatively different behaviour for $\langle \mathbf{s}_0\mathbf{s}_i\rangle$ and $\langle \mathbf{s}_i\mathbf{s}_j\rangle$. While the thermal fluctuations diminish the correlations within the medium at the same time they increase the strength of antiferromagnetic correlations of the central spin. This *order from disorder* phenomenon was observed in several frustrated systems [17, 18, 19, 20] and is connected with a competition between different energy scales. In particular, slightly above $\alpha = 2$ the ground state correlation $\langle \mathbf{s}_0\mathbf{s}_i\rangle_0$ is zero (cf. figure 1), but the fluctuations cause an antiferromagnetic alignment of the spins at finite temperatures.

## 4. The thermodynamic limit

### 4.1. Ground state

Now we turn over to the thermodynamic limit. Then the star represents an antiferromagnet frustrated by an interaction with an extra spin $\mathbf{s}_0$. This situation is somewhat similar to the slightly doped high $T_c$ cuprate superconductors where the holes at the oxygen sites create antiferromagnetically coupled extra-spins frustrating the antiferromagnetic copper matrix [21, 22, 23, 24].

We define the normalized quantum number of the medium

$$x = \frac{2}{N}\, r \quad ; \quad 0 \leq x \leq 1. \tag{13}$$

The ground state energy is obtained by equation (3) by selecting the lowest eigenvalue $E_R$ for a given quantum number $r$ and finding that $r$ which minimizes the energy for given $J_1$ and $J_2$. $E_R$ of the LM antiferromagnet is given by the analytic expression

$$E_R^{LM} = \frac{x^2}{2} - \frac{1}{4}. \tag{14}$$

For the LC no simple analytic expression is available, the $E_R^{LC}(x)$-function was calculated by the numerical solution of the Bethe-Ansatz equations [3, 4]. However, near maximal polarization, $x \to 1$, we can extract from the Bethe-Ansatz an analytic relation for the energy $E_R$

$$E_R^{LC}(x) = \frac{1}{4} - (1-x) + \frac{\pi^2}{48}(1-x)^3 + O((1-x)^5). \tag{15}$$

For the square lattice only approximative results are known. We calculated $E_R^{SL}(x)$ for lattices of $N = 16$, 18, 20 and 24 sites with periodical boundary conditions. The energy scales with $N^{-3/2}$ [25, 26, 27]. By interpolation between the discrete points calculated for $N = 16, 18, 20, 24$ sites and extrapolation to $N \to \infty$ we obtained the numerical data for $E_R^{SL}$ in the thermodynamic limit. The error of these SL data could be estimated by comparing our result for $x = 0$ (antiferromagnetic singlet ground state) with best available results of various methods (variational Monte Carlo, world-line Monte Carlo, spin-wave theories, see the review [28]). We found an error of less then 3%.

A general result (independent of the model and the size of the system) is the stability of the fully polarized ($x = 1$) ferromagnetic state of the embedding medium till precisely $\alpha = \frac{1}{4}$. At this point a second order transition to a canted magnetic structure occurs. Only at infinite $J_2/J_1$ the antiferromagnetic singlet state is reached. In more detail we want to discuss the ground state energy, the spin-spin correlation and the order parameters.

*Energy* – In figure 3 we present the ground state energy in dependence on $\alpha = J_2/J_1$. We find a maximum slightly above $\alpha = \frac{1}{4}$. For the LM star the maximum is at $\alpha = 1/\sqrt{8} \approx 0.35$. For the LC and the SL star the position of the maxima is obtained from the numerical data at 0.26 (LC) and 0.30 (SL). The maximum in $E(x)$ indicates the region of strongest frustration and coincides according to the Hellmann-Feynman [29] theorem with the point where the nearest-neighbour correlation function $\langle \mathbf{s}_i \mathbf{s}_j \rangle_0$ vanishes (maximal spin canting).

*Spin-spin correlation* – In figure 4 the spin-spin correlation functions $\langle \mathbf{s}_0 \mathbf{s}_i \rangle_0$ and $\langle \mathbf{s}_i \mathbf{s}_j \rangle_0$ $(i, j \neq 0,\ j = i+1$ for LC, $j = i + \hat{x}$ or $j = i + \hat{y}$ for SL and $j \in A$, $i \in B$ for LM) are presented. For the LM system these correlation functions are given by explicit formulae

$$\langle \mathbf{s}_0 \mathbf{s}_i \rangle_0 = -\frac{1}{16}\frac{1}{\alpha} \quad ; \quad \alpha \geq \alpha_{crit} \tag{16}$$

$$\langle \mathbf{s}_i \mathbf{s}_j \rangle_0 = \frac{1}{32\alpha^2} - \frac{1}{4} \quad ; \quad \alpha \geq \alpha_{crit} \quad . \tag{17}$$

For the LC and the SL stars they are a result of numerical calculations. It is evident that the physical properties are changed drastically in a small parameter region slightly above the critical $\alpha_{crit} = \frac{1}{4}$. For the LC system it is even suggested by the numerical data that the correlation changes at this point with an infinite slope. This can be checked analytically using the equation (15). We obtain with $\beta = \alpha - \frac{1}{4}$, $\beta \ll 1$

$$\langle \mathbf{s}_0 \mathbf{s}_i \rangle_0 = -\frac{1}{4} + \frac{2}{\pi}\beta^{1/2} - \frac{4}{\pi}\beta^{3/2} + O(\beta^{5/2}) \quad ; \quad \alpha \geq \alpha_{crit} \tag{18}$$

$$\langle \mathbf{s}_i \mathbf{s}_j \rangle_0 = \frac{1}{4} - \frac{8}{\pi}\beta^{1/2} + \frac{80}{3\pi}\beta^{3/2} + O(\beta^{5/2}) \quad ; \quad \alpha \geq \alpha_{crit} \quad . \tag{19}$$

In other words, for the infinite LC star the transition at $\alpha = \frac{1}{4}$ is extremely sharp but still of second order. The LM correlation has for $\beta \to 0$ a finite slope and the correlation behaves according to $\langle \mathbf{s}_0 \mathbf{s}_i \rangle_0 = -\frac{1}{4} + \beta + O(\beta^2)$ and $\langle \mathbf{s}_i \mathbf{s}_j \rangle_0 = \frac{1}{4} - 4\beta + O(\beta^2)$. The behaviour of the SL star is just intermediate between the two limiting cases LC and LM.

*Order parameters* – On the base of the spin correlation we can define the order parameters

$$M^2 = \langle [\frac{1}{N}\sum_{i=1}^{N} \mathbf{s}_i]^2 \rangle = \frac{1}{N^2}\sum_{i,j=1}^{N} \langle \mathbf{s}_i \mathbf{s}_j \rangle \tag{20}$$

for ferromagnetic and

$$M_s^2 = \langle [\frac{1}{N}\sum_{i=1}^{N} \tau_i \mathbf{s}_i]^2 \rangle = \frac{1}{N^2}\sum_{i,j=1}^{N} \tau_i \tau_j \langle \mathbf{s}_i \mathbf{s}_j \rangle \ ; \ \tau_i = +1 \, , \, i \in A \ ; \ \tau_i = -1 \, , \, i \in B \tag{21}$$

for antiferromagnetic long-range order within the medium. We choose here the definitions via the long-range part of the correlation functions and do not introduce symmetry breaking field (cf. [13, 14, 15]). The ferromagnetic order parameter can be expressed directly by the normalized quantum number $x$ defined in (13), $M^2 = \frac{1}{4}x^2$ and can be calculated numerically for LC and SL, but analytically for LM. The antiferromagnetic order parameter can be evaluated for the LM system only. We have

$$M^2_{LM} = \frac{1}{64\alpha^2} \quad ; \quad \alpha \geq \alpha_{crit} \tag{22}$$

$$M^2_{s,LM} = \frac{1}{4} - \frac{1}{64\alpha^2} \quad ; \quad \alpha \geq \alpha_{crit} \quad . \tag{23}$$

For the ferromagnetic order parameter $M^2_{LC}$ of the LC system an analytic expression is available near the second order transition, i.e. for $\beta = \alpha - \frac{1}{4} \ll 1$

$$M^2_{LC} = \frac{1}{4}\left(1 - \frac{16}{\pi}\beta^{1/2} + \frac{64}{\pi^2}\beta + \frac{32}{\pi}\beta^{3/2} + O(\beta^2)\right) \quad ; \quad \alpha \geq \alpha_{crit} \quad . \tag{24}$$

To discuss the magnetic long range order in the whole parameter range we use instead of the ferro- and antiferromagnetic order parameters a parameter

$$P = \frac{1}{N^2}\sum_{i,j=1}^{N} |\langle \mathbf{s}_i \mathbf{s}_j \rangle|. \tag{25}$$

$P$ measures the total pair correlations. For collinear ferro- (i.e. for dominating $J_1$) or collinear antiferromagnetic order (i.e. for dominating $J_2$) $P$ coincides with the corresponding order parameters $M^2$ and $M^2_s$. For the LM system $P$ is given by

$$P = \frac{1}{8} + \left|\frac{1}{8} - \frac{1}{64}\frac{1}{\alpha^2}\right| \quad ; \quad \alpha \geq \alpha_{crit} \quad . \tag{26}$$

Because for LC and SL the order parameter $P$ is not known in the thermodynamic limit we have calculated $P$ with $N = 24$ by exact diagonalization. The results for $P$ are presented in figure 5. In accordance with the behaviour of the spin correlations presented in figure 4 there is a rapid change from the ferromagnetic long-range order for dominating $J_1$ via a canted structure to a state with dominating antiferromagnetic correlations. This antiferromagnetic state is long-range ordered for the LM and SL but possesses no long-range order in LC because of the extremely large quantum fluctuations in one dimension. There is a sharp minimum for LM and SL indicating the area with weak pair correlation. This minimum coincides with the maximum of the energy, i.e. with the point of maximal frustration. For LC the behaviour is qualitatively different, because in contrast to SL and LM the phase for large $\alpha$ is not long-range ordered.

Of course, for SL and LC we have to take into consideration the finite size effects. Due to the contributions of the short range correlations $P$ is overestimated for finite systems, i.e. $P(N) > P(\infty)$. For instance, the value for $P$ larger than $\frac{1}{4}$ at small $\alpha$ and the non-vanishing $P$ for the LC at larger values of $\alpha$ is a consequence of the finite size. Furthermore, we can expect that for SL the minimum in $P$ at strong frustration goes to zero in the thermodynamic limit, i.e. there is no long-range order for $\alpha$ in a region around 0.3.

*4.2. Thermodynamics for the Lieb-Mattis star*

The LM star is distinguished from the other systems by the existence of two additional integrals of motion, namely the square of the sublattice spins $\mathbf{S}_{R_{A(B)}} = \sum_{i\in A(B)} \mathbf{s}_i$ ($\mathbf{S}_R = \mathbf{S}_{R_A} + \mathbf{S}_{R_B}$). This fact allows to find explicitly **all** eigenvalues $E_{s,r,r_A,r_B}$

$$E = \frac{1}{2N} J_1 [s(s+1) - r(r+1) - \frac{3}{4}] + \frac{2}{N^2} J_2 [r(r+1) - r_A(r_A+1) - r_B(r_B+1)] \quad (27)$$

where we have for the relevant quantum numbers

$$r_{A(B)} = 0, 1, \cdots, \frac{N}{4} \qquad \text{for } \mathbf{S}^2_{R_{A(B)}}, \quad (28)$$

$$r = |r_A - r_B|, \cdots, r_A + r_B \quad \text{for } \mathbf{S}^2_R, \quad (29)$$

$$S = |r - \frac{1}{2}|, r + \frac{1}{2} \qquad \text{for } \mathbf{S}^2. \quad (30)$$

The degeneracy to the quantum numbers $r_{A(B)}$ is

$$d_{r_{A(B)}} = \frac{2r_{A(B)} + 1}{\frac{N}{4} + r_{A(B)} + 1} \begin{pmatrix} \frac{N}{2} \\ \frac{N}{4} - r_{A(B)} \end{pmatrix}. \quad (31)$$

In addition we have to take into account the Kramers degeneracy (z-component of the total spin) as $(2s+1)$ to calculate the partition function $Z$

$$Z = e^{\frac{3}{8N}\beta J_1} \sum_{r_A=0}^{N/4} \sum_{r_B=0}^{N/4} d_{r_A} d_{r_B} \exp\left[\beta \frac{2J_2}{N^2} \left[r_A(r_A+1) + r_B(r_B+1)\right]\right]$$
$$\times \sum_{r=|r_A-r_B|}^{r_A+r_B} \exp\left[-\beta \frac{4J_2 - NJ_1}{2N^2} r(r+1)\right] \sum_{s=|r-1/2|}^{r+1/2} (2s+1) \exp\left[-\beta \frac{J_1}{2N} s(s+1)\right] \quad (32)$$

with $\beta = 1/(k_B T)$. In the thermodynamic limit the saddle-point approximation becomes exact, i.e. the sum in (32) is determined by its largest term. Due to symmetry we have $r_A = r_B$. Defining the normalized sublattice polarization

$$y = \frac{4}{N} r_A = \frac{4}{N} r_B, \quad 0 \le y \le 1 \quad (33)$$

we have for the total polarization of the medium $0 \le x \le y$. There exist two phases, which are described by the characteristics of spin correlation and the order parameters of the medium. One is the ferromagnetic phase (F) for dominating $J_1$ ($\alpha < \frac{1}{4}$ and the other one is the canted (or twisted) phase (C) realized only if $\alpha > \frac{1}{4}$. This latter one goes over smoothly in the antiferromagnetic phase for $\alpha \gg 1$. The thermodynamic equations are as follows:

**Case 1:** $J_2 \le \frac{1}{4} J_1$.

Here only the ferromagnetic phase is realized, i.e. the polarizations of the sublattice $A$ and $B$ are parallel and consequently we have $r = r_A + r_B$ or $x = y$. The free energy per site is calculated as

$$F = F_F = -k_B T \mathrm{g}(y) + \frac{1}{4}(J_2 y^2 - J_1 y), \quad (34)$$

with

$$2\mathrm{g}(y) = 2\ln(2) - (1+y)\ln(1+y) - (1-y)\ln(1-y). \tag{35}$$

The sublattice polarization $y$ is determined via a self-consistency equation

$$y = \tanh\left[-\frac{1}{2}\beta J_2 y + \frac{1}{4}\beta J_1\right]. \tag{36}$$

The spin correlations and order parameters are

$$\langle \mathrm{s}_0 \mathrm{s}_i \rangle = -\frac{1}{4}y \quad ; \quad \langle \mathrm{s}_i \mathrm{s}_j \rangle|_{i\in A, j\in B} = \frac{1}{4}y^2 \quad ; \quad \langle \mathrm{s}_i \mathrm{s}_j \rangle|_{i,j\in A(B)} = \frac{1}{4}y^2 \tag{37}$$

$$M^2 = \frac{1}{4}y^2 \; ; \; M_s^2 = 0 \; ; \; P = M^2 = \frac{1}{4}y^2 \quad . \tag{38}$$

**Case 2:** $J_2 > \frac{1}{4}J_1$.
In this case we have two phases separated by a second order transition at a critical temperature $T_G$ (see figure 6). Below $T_G$ we have the canted C phase and above $T_G$ the F phase described by the relations (34-38). In the C phase we have $x = 1/(4\alpha)$ and the free energy reads

$$F = F_C = -k_B T \mathrm{g}(y) - \frac{1}{4}J_2 y^2 - \frac{1}{32}\frac{J_1^2}{J_2} \tag{39}$$

where the sublattice polarization $y$ is determined by

$$y = \tanh\left[\frac{1}{2}\beta J_2 y\right]. \tag{40}$$

For the spin correlation and the order parameters we get

$$\langle \mathrm{s}_0 \mathrm{s}_i \rangle = -\frac{1}{16\alpha} \quad ; \quad \langle \mathrm{s}_i \mathrm{s}_j \rangle|_{i\in A, j\in B} = \frac{1}{32\alpha^2} - \frac{1}{4}y^2 \quad ; \quad \langle \mathrm{s}_i \mathrm{s}_j \rangle|_{i,j\in A(B)} = \frac{1}{4}y^2 \tag{41}$$

$$M^2 = \frac{1}{64\alpha^2} \quad ; \quad M_s^2 = \frac{1}{4}y^2 - \frac{1}{64\alpha^2} \quad ; \quad P = \left|\frac{1}{64\alpha^2} - \frac{1}{8}y^2\right| + \frac{1}{8}y^2 \quad . \tag{42}$$

The critical temperature $T_G$ determined by

$$y(T = T_G) = \frac{1}{4\alpha} \tag{43}$$

is presented in figure 6. For small $T_G$ (that means for $J_2 \to J_1/4$) we have an explicit analytic expression

$$k_B T_G \approx -\frac{J_2}{\ln(\frac{1}{2} - \frac{1}{8\alpha})} \quad . \tag{44}$$

The specific heat and the spin correlations are presented in figures 7 and 8. The second order transition for $\alpha > \frac{1}{4}$ is reflected by the kink in the spins correlations as well as in the peak of the specific heat. The molecular field like character of the LM system in the thermodynamic limit is seen by th shape of the specific heat curve (figure 7). Interesting is the low temperature behaviour for strongly competing $J_1$ and $J_2$. Similar to the finite system we find *order from disorder* behaviour, however, this time in the correlation of the medium $\langle \mathbf{s}_i \mathbf{s}_j \rangle$ and not in $\langle \mathbf{s}_0 \mathbf{s}_i \rangle$ as for the finite system (compare figure 2 and figure 8). In particular, for maximum frustration at $\alpha = 1/\sqrt{8}$ the correlation $\mathbf{s}_i \mathbf{s}_j|_{i \in A, j \in B}$ is completely suppressed for zero temperature and increases with $T$ until the second order transition at $T_G$ (figure 8). We argue that the competition between $J_1$ and $J_2$ is influenced by thermal fluctuations and effectively the $J_1$ coupling starts to overcome the competing $J_2$ at finite $T$. For $\alpha > 1/\sqrt{8}$ the correlation function even changes its sign with increasing temperature.

## 5. Conclusions

In conclusion, we present the Heisenberg star with competing interactions (frustration) as an example for a solvable quantum spin system. The solution can be given whenever the solution of the surrounding medium is known. We consider as embedding media the linear chain, the square lattice (where only an approximative solution is given) as well as the Lieb-Mattis antiferromagnet which represent systems with different strengths of the quantum fluctuations. The effect of competing interactions manifests itself in a maximum of the ground state energy at maximal frustration and in a weakening of the magnetic correlation. In the region of maximal frustration there is a rapid change of correlation functions when the strength of competition is varied. This change of correlation is in particular dramatic when the quantum fluctuations are strong (LC). The competition between the exchange interactions may yield at finite temperatures *order from disorder* phenomena, i.e. the strength of magnetic correlation can be increased by thermal fluctuations.

## Acknowledgments

This work was supported by the Deutsche Forschungsgemeinschaft (project Ri 615/1-2).

## Figure captions

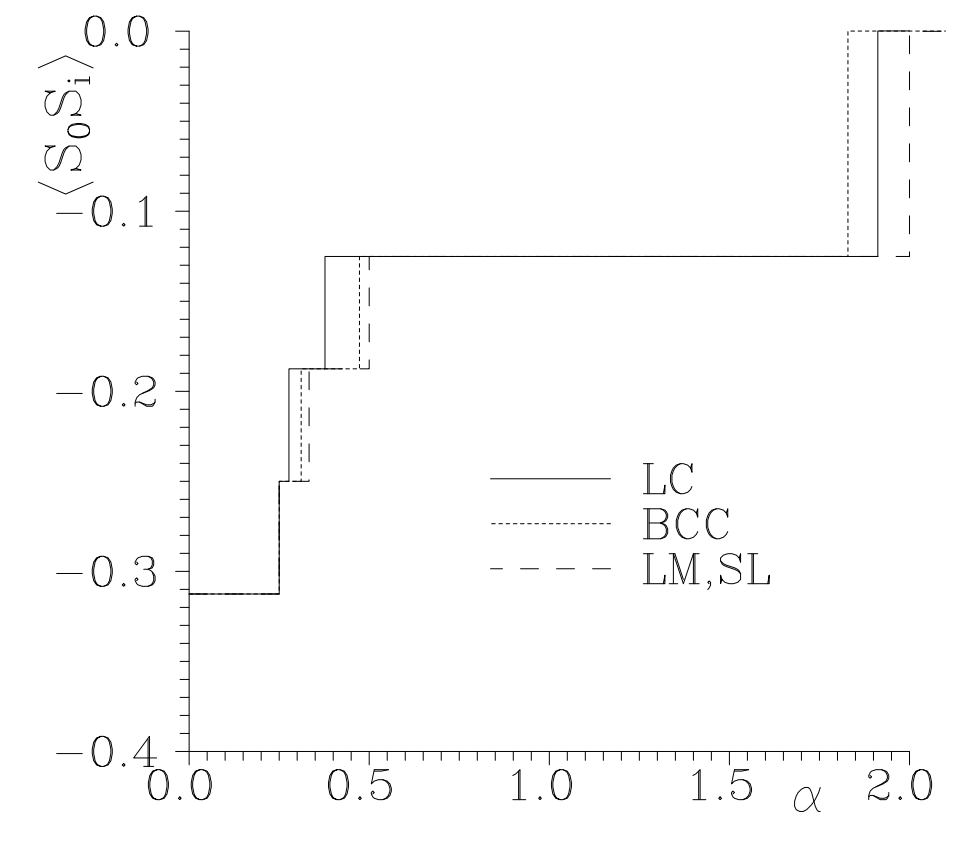

**Figure 1.** Ground state spin correlation $\langle \mathbf{s}_0 \mathbf{s}_i \rangle$ versus $\alpha = J_2/J_1$ for a finite cluster of a central spin $\mathbf{s}_0$ with $N = 8$ neighbouring spins $\mathbf{s}_i$ for various arrangements of neighbouring spins (LC - linear chain, BCC - elementary unit of a body centered lattice, LM - Lieb-Mattis, SL - square lattice).

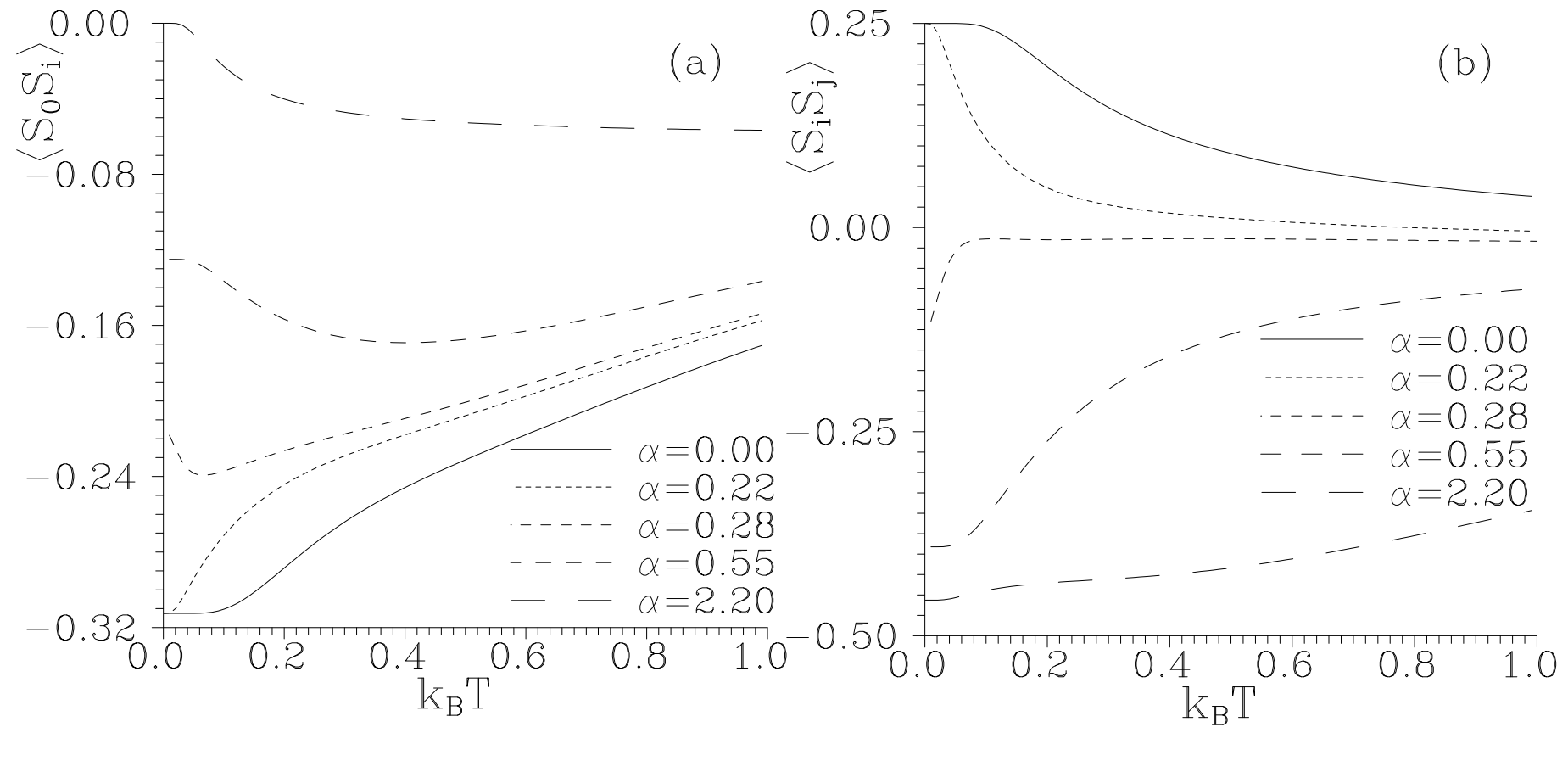

**Figure 2.** Spin correlations $\langle \mathbf{s}_0 \mathbf{s}_i \rangle$ (a) and $\langle \mathbf{s}_i \mathbf{s}_j \rangle$ ($i$ and $j$ are neighbouring sites) (b) versus temperature $T$ at various $\alpha = J_2/J_1$ for a finite cluster of a central spin $\mathbf{s}_0$ with $N = 8$ neighbouring spins $\mathbf{s}_i$ which form a linear chain (LC).

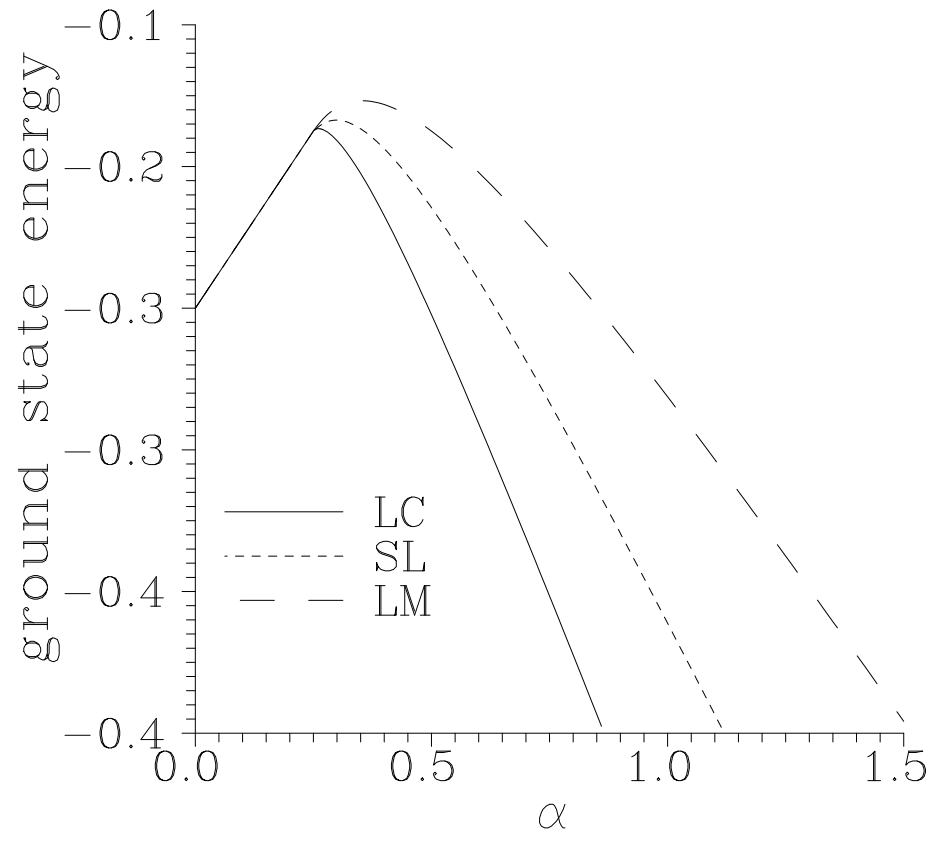

**Figure 3.** Ground state energy versus frustration parameter $\alpha$ for $N \to \infty$.

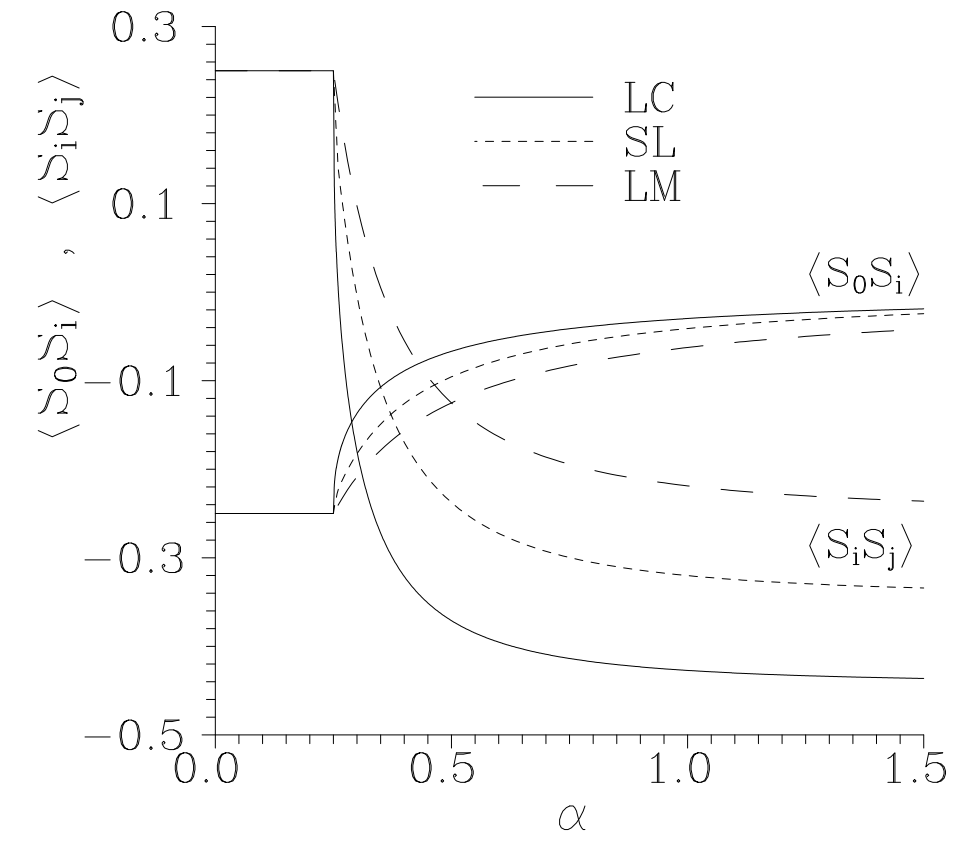

**Figure 4.** Ground-state spin correlation within the medium $\langle \mathbf{s}_i \mathbf{s}_j \rangle$ and between the central spin and the embedding medium $\langle \mathbf{s}_0 \mathbf{s}_i \rangle$ versus $\alpha$ in the thermodynamic limit $N \to \infty$.

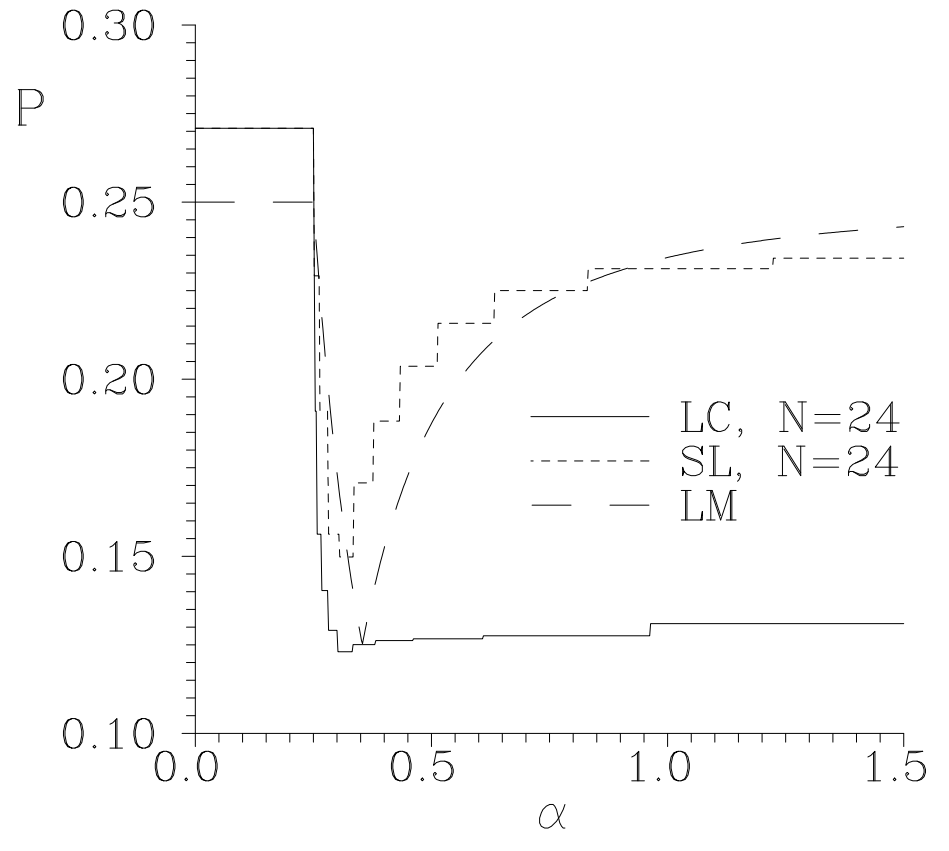

**Figure 5.** Ground-state total correlation parameter $P$ (cf. equation (25)) of the medium versus $\alpha$.

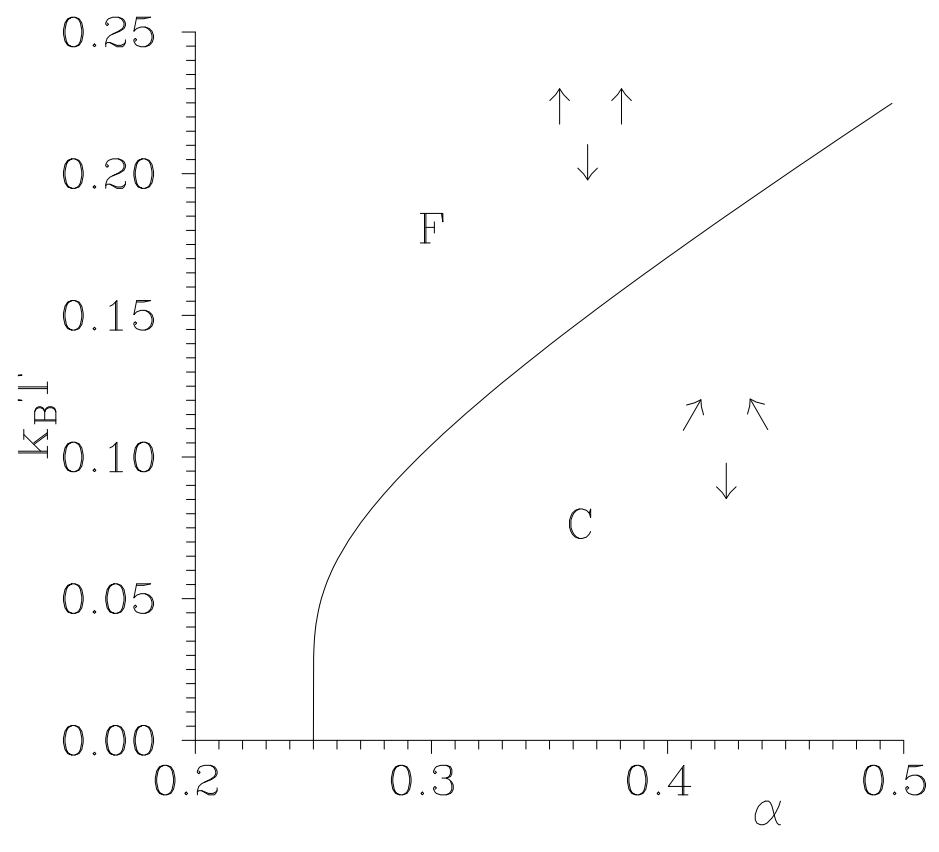

**Figure 6.** Transition temperature $T_G$ versus $\alpha$ for the LM system and illustration of the principal arrangement of the spins in the ferromagnetic F phase and the canted C phase. The lower arrow represents the central spin and the two upper arrows represent the sublattice spins of the LM system.

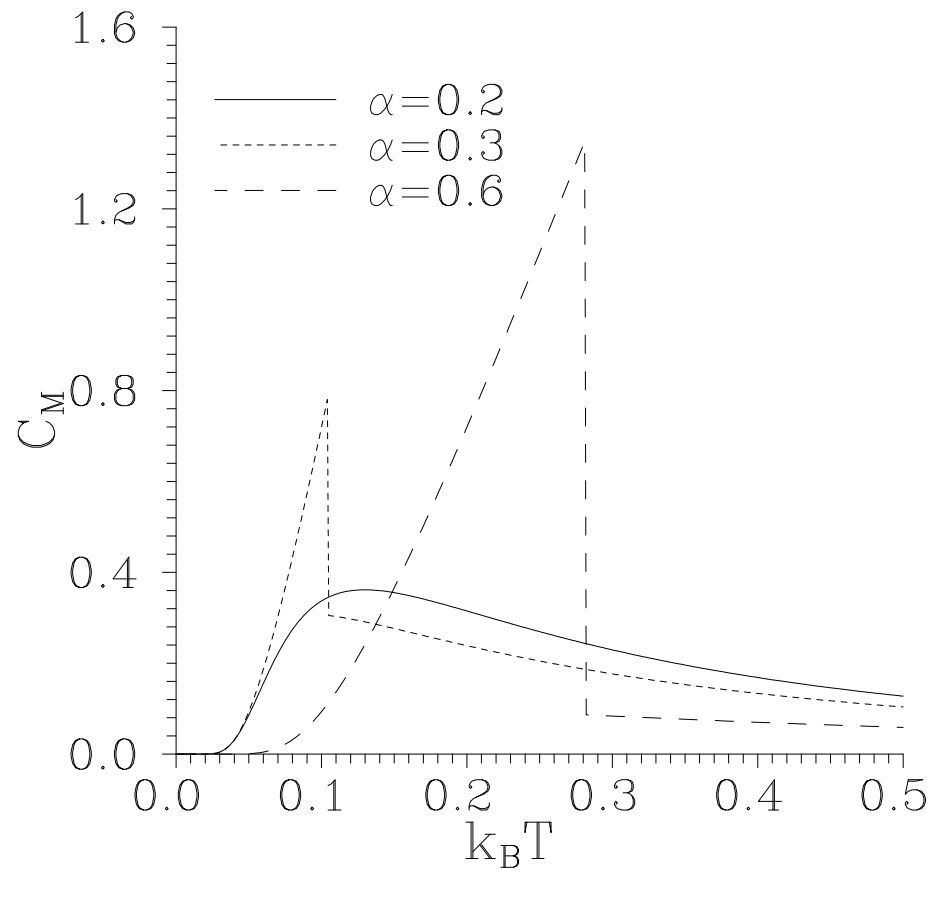

**Figure 7.** Specific heat of the LM star versus temperature $T$ for several $\alpha$.

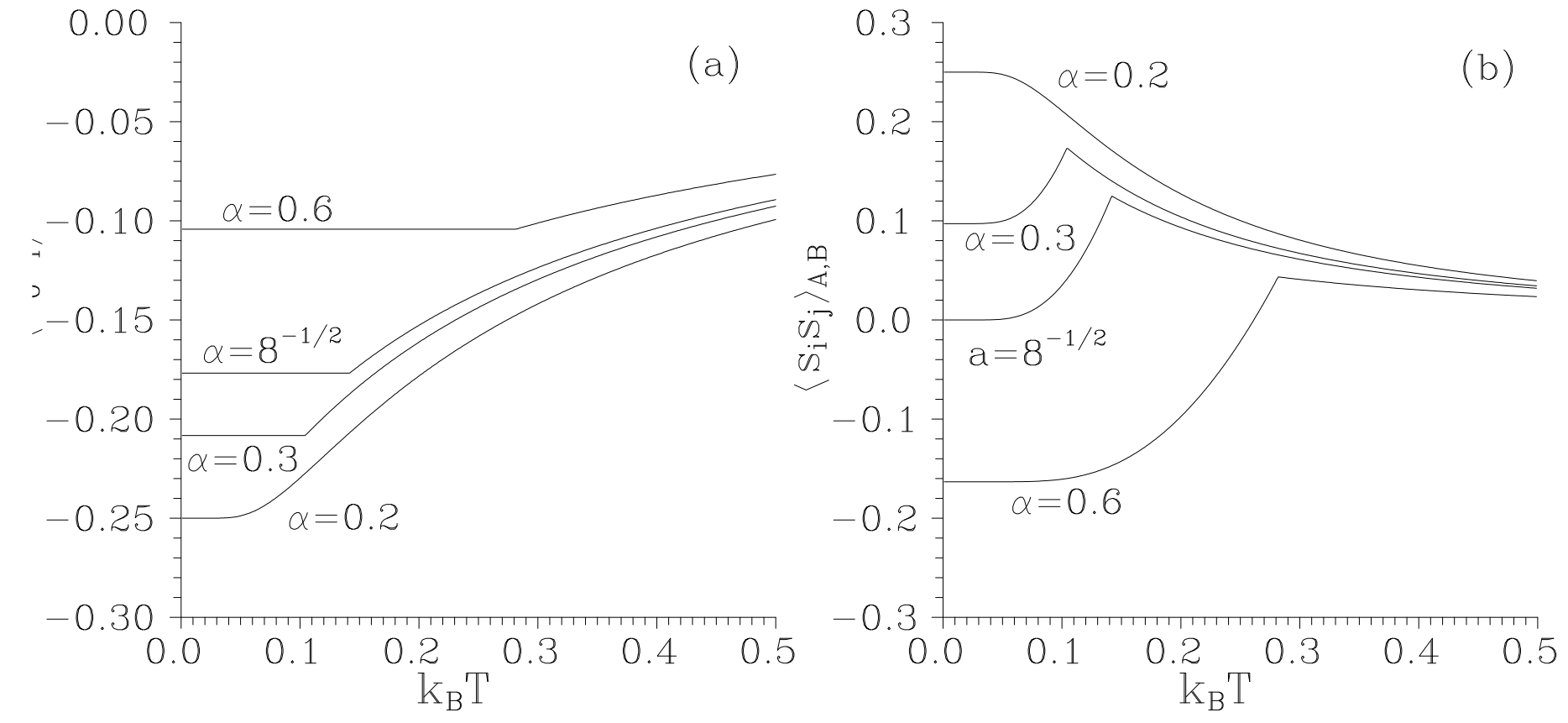

**Figure 8.** Spin correlation of the LM star between the central spin and the embedding medium $\langle \mathbf{s}_0 \mathbf{s}_i \rangle$ (a) and within the medium $\langle \mathbf{s}_i \mathbf{s}_j \rangle$ (b) versus temperature $T$ for several $\alpha$.